# Comparing human and automatic thesaurus mapping approaches in the agricultural domain

Boris Lauser
Gudrun Johannsen
Caterina Caracciolo
Johannes Keizer
FAO,
Italy
boris.lauser@fao.org
gudrun.johannsen@fao.org
caterina.caracciolo@fao.org
johannes.keizer@fao.org

Willem Robert van Hage
TNO Science & Industry /
Vrije Universiteit
Amsterdam,
the Netherlands
wrvhage@few.vu.nl

Philipp Mayr
GESIS Social Science
Information Centre
Bonn,
Germany
philipp.mayr@gesis.org

## Abstract

Knowledge organization systems (KOS), like thesauri and other controlled vocabularies, are used to provide subject access to information systems across the web. Due to the heterogeneity of these systems, mapping between vocabularies becomes crucial for retrieving relevant information. However, mapping thesauri is a laborious task, and thus big efforts are being made to automate the mapping process. This paper examines two mapping approaches involving the agricultural thesaurus AGROVOC, one machine-created and one human created. We are addressing the basic question "What are the pros and cons of human and automatic mapping and how can they complement each other?" By pointing out the difficulties in specific cases or groups of cases and grouping the sample into simple and difficult types of mappings, we show the limitations of current automatic methods and come up with some basic recommendations on what approach to use when.

## 1. Introduction

Information on the Internet is constantly growing and with it the number of digital libraries, databases and information management systems. Each system uses different ways of describing their metadata, and different sets of keywords, thesauri and other knowledge organization systems (KOS) to describe its subject content. Accessing and synthesizing information by subject across distributed databases is a challenging task, and retrieving all information available on a specific subject in different information systems is nearly impossible. One of the reasons is the different vocabularies used for subject indexing. For example, one system might use the keyword 'snakes', whereas the other system uses the taxonomic name 'Serpentes' to classify information about the same subject. If users are not aware of the different 'languages' used by the systems, they might not be able to find all the relevant information. If, however, the system itself "knows", by means of mappings, that 'snakes' maps to 'Serpentes', the system can appropriately translate the user's query and therefore retrieve the relevant information without the user having to know about all synonyms or variants used in the different databases.

Mapping major thesauri and other knowledge organization systems in specific domains of interest can therefore greatly enhance the access to information in these domains. System developers for library search applications can programmatically incorporate mapping files into the search applications. The mappings can hence be utilized at query time to translate a user query into the terminology used in the different systems of the available mappings and seamlessly retrieve consolidated information from various databases[1].

Mappings are usually established by domain experts, but this is a very labor intensive, time consuming and error-prone task (Doerr, 2001). For this reason, great attention is being devoted to the possibility of creating mappings in an automatic or semi-automatic way (Vizine-Goetz, Hickey, Houghton, Thompsen, 2004), (Euzenat & Shvaiko, 2007), (Kalfoglou & Schorlemmer, 2003) and (Maedche, Motik, Silva, Volz, 2002). However, so far, research has focused mainly on the quantitative analysis of the automatically obtained mappings, i.e. purely in terms of precision and recall of either end-to-end document retrieval or of the quality of the sets of mappings produced by a system. Only little attention has been paid to a comparative study of manual and automatic mappings. A qualitative analysis is necessary to learn how and when automatic techniques are a suitable alternative to high-quality but very expensive manual mapping. This paper aims to fill that gap. We will elaborate on mappings between three KOS in the agricultural domain: AGROVOC, NALT and SWD.

- AGROVOC[2] is a multilingual, structured and controlled vocabulary designed to cover the terminology of all subject fields in agriculture, forestry, fisheries, food and related domains (e.g. environment). The AGROVOC Thesaurus was developed by the Food and Agriculture Organization of the United Nations (FAO) and the European Commission, in the early 1980s. It is currently available online in 17 languages (more are under development) and contains 28,718 descriptors and 10,928 non-descriptors in the English version.

- The NAL Thesaurus[3] (NALT) is a thesaurus developed by the National Agricultural Library (NAL) of the United States Department of Agriculture and was first released in 2002. It contains 42,326 descriptors and 25,985 non-descriptors organized into 17 subject categories and is currently available in two languages (English and Spanish). Its scope is very similar to that of AGROVOC. Some areas such as economical and social aspects of rural economies are described in more detail.

- The Schlagwortnormdatei[4] (SWD) is a subject authority file maintained by the German National Library and cooperating libraries. Its scope is that of a universal vocabulary. The SWD contains around 650,000 keywords and 160,000 relations between terms. The controlled terms cover all disciplines and are classified within 36 subject categories. The agricultural part of the SWD contains around 5,350 terms.

These controlled vocabularies (AGROVOC, NALT, and SWD) have been part of two mapping initiatives, conducted by the Ontology Alignment Evaluation Initiative (OAEI) and by the GESIS Social Science Information Centre (GESIS-IZ) in Bonn.

The **Ontology Alignment Evaluation Initiative (OAEI)** is an internationally-coordinated initiative to form consensus on the evaluation of ontology mapping methods. The goal of the OAEI is to help to improve the work on ontology mapping by organizing an annual comparative evaluation of ontology mapping systems on various tasks. In 2006 and 2007 there was a task that consisted in mapping the AGROVOC and NALT thesauri, called the "food task." A total of eight systems participated in this event. For this paper we consider the results of the five best performing systems that participated in the OAEI 2007 food task: Falcon-AO, RiMOM, X-SOM, DSSim and SCARLET. Details about this task, the data sets used and the results obtained can be found on the website of the food task[5]. The mapping relations that participants could use were the

---

SKOS Mapping Vocabulary relations exactMatch, broadMatch, and narrowMatch, because these correspond to the thesaurus constructs most people agree on: USE, BT and NT.

In 2004, the German Federal Ministry for Education and Research funded a major terminology mapping initiative called **Competence Center Modeling and Treatment of Semantic Heterogeneity**[6] at the GESIS-IZ, which published its conclusion at the end of 2007 (see project report in Mayr & Petras, 2008a, to be published). The goal of this initiative was to organize, create and manage mappings between major controlled vocabularies (thesauri, classification systems, subject heading lists), initially centred around the social sciences but quickly extending to other subject areas. To date, 25 controlled vocabularies from 11 disciplines have been intellectually (manually) connected with vocabulary sizes ranging from 1,000-17,000 terms per vocabulary. More than 513,000 relations were constructed in 64 crosswalks. All terminology-mapping data is made available for research purposes. We also plan on using the mappings for user assistance during initial search query formulation as well as for ranking of retrieval results (Mayr, Mutschke, Petras, 2008). The evaluation of the value added by mappings and the results of an information retrieval experiment using human generated terminology mappings is described in (Mayr & Petras, 2008b, to be published). The AGROVOC – SWD mapping was created within this initiative in 2007.

## 2. Related Work

Many thesauri, amongst which AGROVOC and the Aquatic Sciences and Fisheries Abstracts Thesaurus (ASFA)[7] are being converted into ontologies, in order to enhance their expressiveness and take advantage of the tools made available by the semantic web community. Therefore, great attention is being dedicated also to mapping ontologies. An example is the Networked Ontologies project (NeOn)[8], where mappings are one of the ways to connect ontologies in networks.

Within NeOn, an experiment was carried out to automatically find mappings between AGROVOC and ASFA. Since ASFA is a specialized thesaurus in the area of fisheries and aquaculture, the mapping with AGROVOC resulted in a mapping with the fisheries-related terms of AGROVOC. The mappings were extracted by means of the SCARLET system (cf. section 3) and were of three types: superclass/subclass, disjointness and equivalence. Evaluation was carried out manually by two FAO experts, in two runs: first with a sample of 200 randomly selected mappings, then with a second sample of 500 mappings. The experts were also supported in their evaluation by the graphical interface. The results obtained were rather poor (precision was 0.16 in the first run of the evaluation and 0.28 in the second run), especially if compared with the high results obtained by the same system with the mapping of AGROVOC and NALT (cf. section 3). The hypothesis formulated to explain this low performance is related to the low degree of overlap between AGROVOC and ASFA,[9] and that the terms in ASFA may not be well covered by the Semantic Web, as required by SCARLET. Cases like this clearly show how beneficial it would be to gain a clear understanding of when manual mapping is more advisable than automatic mapping (as in the case of the AGROVOC- ASFA mapping) or the other way around (as in the case of the AGROVOC - NALT mapping analyzed in this paper).

Another major mapping exercise was carried out mapping AGROVOC to the Chinese Agricultural Thesaurus (CAT) described in (Liang et al., 2006). The mapping has been carried out using the SKOS Mapping Vocabulary[10] (version 2004) and addresses another very important issue in mapping thesauri and other KOS: multilinguality. AGROVOC has been translated from

---

[6] The project was funded by BMBF, grant no. 01C5953.
http://www.gesis.org/en/research/information_technology/komohe.htm.
[7] http://www4.fao.org/asfa/asfa.htm.
[8] http://neon-project.org.
[9] In particular, a problem could be the different level of details of the two resources, as ASFA tends to be very specific on fisheries related terms.
[10] http://www.w3.org/2004/02/skos/mapping/spec/.

English to Chinese, whereas CAT has been translated from Chinese to English. This creates potential problems as the following example illustrates: CAT '水稻'/'Oryza sativa' was originally mapped to AGROVOC 'Oryza sativa'. However, upon closer examination, the Chinese lexicalization in AGROVOC of 'Oryza sativa', which is '稻', appears to be the broader term of the CAT Chinese term. Moreover, a search in AGROVOC for the CAT Chinese term '水稻', shows the English translation as 'Paddy'. These discrepancies indicate the weakness of the above mentioned procedure and the necessity of cross checking all lexicalizations in both languages. Such cases pose hard problems for automatic mapping algorithms and can only be addressed with human support at the moment.

Other related work on semantic interoperability can be found in (Patel et al., 2005).

## 3. The AGROVOC – NALT mapping within the OAEI

In the OAEI 2007 food task, five systems using distinct mapping techniques were compared on the basis of manual sample evaluation. Samples were drawn from each of the sets of mappings supplied by the systems to measure precision. Also, a number of small parts of the mapping were constructed manually to measure recall. Details about the procedure can be found in (Euzenat et al., 2007). Each participant documented their mapping method in a paper in the Ontology Matching 2007 workshop[11] (Hu et al., 2007), (Li, Zhong, Li, Tang, 2007), (Curino, Orsi, Tanca, 2007), (Nagy, Vargas-Vera, Motta, 2007) and (Sabou, Gracia, Angeletou, d'Aquin, Motta, 2007). Table 1 summarizes, for each system, the type of mapping found, how many mappings were identified and the precision and recall scores measured on the set of returned mappings, where:

Precision = | found mappings ∩ correct mappings | / | found mappings | , and

Recall = | found mappings ∩ correct mappings | / | correct mappings |.

| System | Falcon-AO | RiMOM | X-SOM | DSSim | SCARLET | | |
|---|---|---|---|---|---|---|---|
| **Mapping type** | = | = | = | = | = | < > | null(0) |
| **# mappings** | 15300 | 18419 | 6583 | 14962 | 81 | 6038 | 647 |
| **Precision** | **0.84** | 0.62 | 0.45 | 0.49 | 0.66 | 0.25 | |
| **Recall** | **0.49** | 0.42 | 0.06 | 0.20 | 0.00 | 0.00 | |

Table 1. The OEAI 2007 food task. Results (in terms of precision and recall) of the 5 systems participating in the initiative. Best scores are in **boldface**. All systems found equivalence mappings only, except SCARLET that also found hierarchical mappings.

The system that performed best at the OAEI 2007 food task was Falcon-AO. It found around 80% of all equivalence relations using lexical matching techniques. However, it was unable to find any hierarchical relations. Also, it did not find relations that required background knowledge to discover. This led to a recall score of around 50%. The SCARLET system was the only system that found hierarchical relations using the semantic web search engine Watson[12] (Sabou et al., 2007). Many of the mappings returned by SCARLET were objectively speaking valid, but more generic than any human would suggest. This led to a very low recall score.

## 4. The AGROVOC – SWD mapping in the GESIS-IZ approach

The GESIS-IZ approach considers intellectually (manually) created relations that determine equivalence, hierarchy (i.e. broader or narrower terms), and association mappings between terms from two controlled vocabularies. Typically, vocabularies will be related bilaterally, that means there is a mapping relating terms from vocabulary A (start terms in Table 2) to vocabulary B (end terms) as well as a mapping relating terms from vocabulary B to vocabulary A. Bilateral relations are not necessarily symmetrical. E.g. the term 'Computer' in system A is mapped to term

---

[11] http://www.om2007.ontologymatching.org/
[12] http://watson.kmi.open.ac.uk/

'Information System' in system B, but the same term 'Information System' in system B is mapped to another term 'Data base' in system A. Bilateral mappings are only one approach to treat semantic heterogeneity; compare (Hellweg et al., 2001) and (Zeng & Chan, 2004). The approach allows the following 1:1 or 1:n mappings: Equivalence (=) means identity, synonym, quasi-synonym; Broader terms (<) from a narrower to a broad; Narrower terms (>) from a broad to a narrower; Association (^): mapping between related terms; and Null (0) which means that a term can not be mapped to another term. The first three of these relations correspond to the exactMatch, broadMatch, and narrowMatch relations from the SKOS Mapping Vocabulary.

The AGROVOC-SWD mapping is a fully human generated bilateral mapping that involves major parts of the vocabularies (see Table 2). Both vocabularies were analysed in terms of topical and syntactical overlap before the mapping started. All mappings in the GESIS-IZ approach are established by researchers, terminology experts, domain experts, and postgraduates. Essential for a successful mapping is the complete understanding of the meaning and semantics of the terms and the intensive use of the internal relations of the vocabularies concerned. This includes performing lots of simple syntactic checks of word stems but also semantic knowledge, i.e. to lookup synonyms and other related or associated terms.

| Mapping direction | # mappings | = | < | > | ^ | null 0 | start terms | end terms |
|---|---|---|---|---|---|---|---|---|
| AGROVOC - SWD | 6254 | 5500 (4557 identical) | 100 | 314 | 337 | 3 | 6119 | 6062 |
| SWD - AGROVOC | 11189 | 6462 (4454 identical) | 3202 | 145 | 1188 | 192 | 10254 | 6171 |

Table 2. Mapping of Agrovoc – SWD. Numbers of established mappings by type and by direction.

The establishment of mappings is based on the following practical rules and guidelines:

1) During the mapping of the terms, all existing intra-thesaurus relations (including scope notes) have to be used.
2) The utility of the established relations has to be checked. This is especially important for combinations of terms (1:n relations).
3) 1:1 relations are prior.
4) Word groups and relevance adjustments have to be made consistently.

In the end the semantics of the mappings are reviewed by experts and samples are empirically tested for document recall and precision (classical information retrieval definition). Some examples of the rules in the KoMoHe approach can be found in (Mayr & Petras, 2008a, to be published).

## 5. Qualitative Assessment

Given these two approaches, one completely carried out by human subject experts and the other by machines trying to simulate the human task, the basic questions are: who performs more efficiently in a certain domain?, what are the differences?, and where are the limits? In order to draw some conclusions, a qualitative assessment is needed.

### 5.1 Alignment of the mappings

We first "aligned" the mappings for the overlapping AGROVOC terms that have been mapped both to NALT and to SWD. For this we aligned the AGROVOC term with the mapped NALT terms (in English) and the mapped SWD term (in German): about 5,000 AGROVOC terms have been mapped in both approaches. For the AGROVOC-NALT mapping, we took the entire set of suggestions made by five systems participating in OAEI 2007. We also listed the number of systems that have suggested the mapping between the AGROVOC and the NALT term (between 1 and 5) and the specific mapping that has been assigned in the SWD mapping (equality, broader, narrower or related match). In case of several suggestions for a mapping (For example the AGROVOC term 'Energy value' has been suggested to be mapped to 'energy' as well as 'digestible protein' in the NAL thesaurus; the latter being an obvious mistake made by one of the systems.) we left all the multiple suggestions to be evaluated later.

We then grouped the aligned mappings into the higher level subject categories of AGROVOC and finally into four major terminology groups: Taxonomic, Biological/Chemical, Geographic, and Miscellaneous. These categories are the same as those used in the OAEI food task evaluation.

This was done in order to be able to draw more detailed conclusions on the difficulty of mappings based on the terminology group a particular mapping falls into. These groups were chosen in order to be more specific on whom to contact to evaluate the respective mappings. This will give an indication on what kind of knowledge is generally harder for automatic computer systems to map and what kind of background knowledge might also be needed to solve the more difficult cases.

## 5.2 Rating of a representative sample

Out of the about 5,000 mappings, we chose a representative sample of 644 mappings to be manually assessed. The mappings for the sample have been picked systematically in such a way that each of the groups is represented. We then assigned one of the following 6 difficulty ratings once for each of the mappings, AGROVOC-NALT and AGROVOC-SWD respectively. The assessments were done by Gudrun Johannsen and Willem Robert van Hage. Table 3 summarizes our rating.

| Rating | Explanation |
|---|---|
| 1. Simple | the prefLabels are literally the same / exact match |
| 2. Alt Label | there is a literal match with an alternative label / synonym in the other thesaurus |
| 3. Easy Lexical | the labels are so close that any laymen can see that they are the same terms/concepts |
| 4. Hard Lexical | the labels are very close, but one would have to know a little about the naming scheme used in the thesaurus (e.g. some medical phrases have a different meaning when the order of the words is changed and doctors know that) |
| 5. Easy Background Knowledge | there are no clues as in point 1-4 for a match, but the average adult laymen knows enough to conclude that there is a mapping |
| 6. Hard Background Knowledge | there are no clues as in point 1-4 for a match and you have to be an expert in some field, e.g. agriculture, chemistry, or medicine, to deduce that there is a mapping |

Table 3. Scale used to rate the mapping based on their "difficulty." The scale goes from 1 (Simple) to 6 (Hard Background Knowledge).

## 5.3 Analysis of Examples

The assessment of the sample selection of 644 mappings is summarized in Table 4. The table is grouped by major subject groups: Taxonomic, Biological/Chemical and Miscellaneous. For each mapping approach (AGROVOC-NALT and AGROVOC-SWD), the table shows, what percentage of the mappings in the respective group are Simple, Easy Lexical, etc. The numbers in brackets are the absolute numbers. For example in the group Miscellaneous: 18.12% of the AGROVOC- SWD mappings in this subject group have been found to be of difficulty 6 (Hard Background Knowledge), whereas only 1.45% of the AGROVOC-NALT mappings have been given this rating.

Table 5 shows the mappings that have been wrongly assigned with the automatic approach in the AGROVOC-NALT mapping. In the assessment, we have specified if these wrong mappings should have been broader mappings (>), narrower mappings (<), related term mappings (^) or simply completely wrong, i.e. null (0) and should not have been suggested.

The Geographic group has been left out from the table, since the sample contained only very few mappings (less than 20). In any case, we can make the rather trivial statement that the Geographic group turns out to be rather simple, i.e. there seems to be an overall consensus on country names and other geographic concepts (in our case, the geographic group consists basically of country names). However, we have to be careful with this statement, especially when it comes to geopolitics. Borders of countries and similarly sensitive concepts might be called the same in two systems (and therefore seem simple and would be suggested by an automatic mapping tool with high security), but actually defined differently and mapping the two could raise sensitive issues. Take for example 'Taiwan': In AGROVOC, the non-preferred term 'China (Taiwan)' refers to the preferred term 'Taiwan', which has the broader term (BT) 'China', whereas in NALT 'Taiwan' uf 'China (Taiwan)' has the broader term 'East Asia'. Another example, which is currently an issue, is the concept 'Macedonia'. It has been used in the Codex Alimentarius[13] to refer to the former Yugoslavian Republic of Macedonia. However, since there is also a region in Greece, which is called Macedonia, the Greek authorities have requested the Codex Alimentarius to use 'The former Yugoslavian Republic of' in the name of the concept. Moreover, country definitions are very time dependent. How a user might best map geographical terms depends on the use case. For some purposes automatic mapping is a quick and good solution. For others it might be better to map all geographical terms manually, which is generally feasible due to the relatively small number of countries in the world (as compared, for example, to plant species).

| Taxonomic | Simple | Alt Label | Easy Lexical | Easy Background | Hard Lexical | Hard Background |
|---|---|---|---|---|---|---|
| AG - SWD | 26.82%(70) | 39.08%(102) | 6.90%(18) | 3.45%(9) | 6.51%(17) | 17.24%(45) |
| AG - NALT | 65.13%(170) | 22.61%(59) | 1.15%(3) | 0.00%(0) | 1.92%(5) | 0.00%(0) |
| | | | | | | |
| Biological /Chemical | Simple | Alt Label | Easy Lexical | Easy Background | Hard Lexical | Hard Background |
| AG - SWD | 62.35%(53) | 21.18%(18) | 1.18%(1) | 2.35%(2) | 1.18%(1) | 11.76%(10) |
| AG - NALT | 64.71%(55) | 12.94%(11) | 3.53%(3) | 0.00%(0) | 3.53%(3) | 1.18%(1) |
| | | | | | | |
| Miscellaneous | Simple | Alt Label | Easy Lexical | Easy Background | Hard Lexical | Hard Background |
| AG - SWD | 33.33%(92) | 11.96%(33) | 10.14%(28) | 16.67%(46) | 9.78%(27) | 18.12%(50) |
| AG - NALT | 49.28%(136) | 24.28%(67) | 3.99%(11) | 0.36%(1) | 1.81%(5) | 1.45%(4) |

Table 4. Rating of the mappings by terminology groups (taxonomic, biological, miscellaneous) and by rating of difficulty.

| | should be < | should be > | should be null (0) | should be ^ |
|---|---|---|---|---|
| Taxonomic | 2.68%(7) | 0.38%(1) | 5.75%(15) | 0.38%(1) |
| Biological / Chemical | 2.35%(2) | 1.18%(1) | 10.59%(9) | 0.00%(0) |
| Miscellaneous | 1.45%(4) | 0.36%(1) | 13.77%(38) | 3.26%(9) |

Table 5. Mapping of AGROVOC-NALT. Classification of wrong mappings.

---

[13] The Codex Alimentarius Commission was created in 1963 by FAO and WHO to develop food standards, guidelines and related texts such as codes of practice under the Joint FAO/WHO Food Standards Programme. The main purposes of this Programme are protecting health of the consumers, ensuring fair trade practices in the food trade, and promoting coordination of all food standards work undertaken by international governmental and non-governmental organizations. It is available at: http://www.codexalimentarius.net/web/index_en.jsp.

Analyzing the other groups listed in the table leads to the few first statements: First of all, we can say that in general, Biological/Chemical like Geographical terminology is fairly easy to map (over 60% rated as Simple). This result makes sense, since like for geographical concepts there is probably a good consensus in the world on names of biological entities and chemicals[14]. Taking into account the alternative labels, this statement also holds for the group of taxonomic terminology mapping. Apparently, in the German language there are more discrepancies on the usage of preferred versus non-preferred labels and synonyms than in the English language. The Miscellaneous group (including the majority of mappings) appears to be the most difficult one: 13.77% of the automatically suggested mappings were even wrong, and it shows the highest percentage of Hard Background Knowledge mappings.

Further analyzing the mappings, we found that the AGROVOC-SWD mapping has a considerable amount of broader (>) and narrower (<) mappings. These are in general more difficult to find than equivalence mappings (either very easy or very difficult, because Hard Background Knowledge may be required), and therefore pose a big problem to automatic mapping algorithms. The SWD part on agriculture is also considerably smaller than the AGROVOC or NAL thesaurus and therefore many broader and narrower mappings are possible. Automatic mapping approaches have difficulty with such discrepancies. Apparently, subterms are often a good lexical clue for a < or > relation, but how does a computer decide which of the subterms is the superclass? Sometimes it is easy because one of the subterms is an adjective, while the other is a noun (e.g. 'mechanical damage' is a damage), but sometimes both are nouns (e.g. 'Bos taurus' is a Bos, not a taurus, but 'fruit harvester' is a harvester), and this is hard to parse. There are also cases where lexical inclusion can bring confusion, for example 'Meerrettich' (horseradish is *Armoracia rusticana*) and 'Meerrettichbaum' (horseradish tree is *Moringa oleifera*), as they refer to completely different concepts. Eventually, this problem might be solved by machine learning, but current mapping systems do not have any functionality to detect various common naming conventions.

It is remarkable that for the harder mappings (Hard Lexical, Easy Background, Hard Background), the percentage that has been found by the automatic approaches is overall very little (at most 3.53% for Hard Lexical biological/chemical terms), whereas the manual mapping approach can obviously identify these mappings. For example in the Miscellaneous group, more than 40% of the manual AGROVOC-SWD mappings fall into one of the three hardest ratings. The automatic mappings with this rating accumulate to less than 4%. Table 5 shows the numbers of wrong automatic mapping suggestions. The percentages in the three hardest ratings of the AGROVOC-NALT mapping are obviously cases of wrong suggestions, as listed in Table 5, which were either completely wrong mappings or should have been broader, narrower or related mappings.

It is not impossible, however, for automatic algorithms to also detect even Hard Background Knowledge mappings, for example by means of text mining. Some of these are easier to solve than others, because some background knowledge is simply easier to find. For instance, there are many web pages about taxonomy, but few about 'Lebensmittelanalyse' (food analysis). There are also many about chemicals, but few that state that a 'Heckstapler' (rear stapler) is some kind of 'Handhabungsgeraet' (handling equipment).

Some more concrete examples of mappings of varying difficulty:

1) *Mapping rated Alt label.* AGROVOC-NALT 'Marketing Strategies' = 'Marketing Techniques'. This mapping has been rated 'alt label', since, for example, in AGROVOC, 'Marketing Strategy' is the non-descriptor of 'Marketing Techniques'. This case makes it easy for an automatic classifier. However, this might also be misleading. In the agriculture domain, it might be correct to declare equivalence between these terms. However, in another domain there might actually be no mapping or at most a related term mapping. For example, in the business area, marketing strategies differ from marketing techniques substantially in that the strategies are long term objectives and roadmaps whereas the marketing techniques are operational techniques used in the marketing of certain products. For an automatic mapping algorithm, this is difficult to detect and alternative labels as they are sometimes found in thesauri, might be misleading.

---

[14] Organizations like The American Chemical Society (CAS, http://www.cas.org/expertise/cascontent/registry/) maintains lists of unique identifiers for chemicals in various languages. Various resources are also available that relate various chemical names to their CAS identifiers.

2) *Mapping rated Hard Background Knowledge.* Both in AGROVOC and the NAL Thesaurus there is the term 'falcons' (exact match, simple mapping) while in SWD the German term 'Falke' does not exist, and thus had to be mapped to the broader term 'Greifvoegel' (predatory birds) which requires human background knowledge. However, in this case, the human knowledge could be found by a mapping system, if it would exploit the German Wikipedia. On the page about Falke[15], it states: "Die Falken (Gattung Falco) sind Greifvögel...".

3) *Mapping rated Hard Background Knowledge.* In SWD the term 'Laubfresser' (folivore) which does not exist in AGROVOC or in NALT had to be mapped to the broader term 'Herbivore'. This is another example where Hard Background Knowledge is needed.

4) Sometimes terms which seem to match exactly are incorrectly machine-mapped, for example when they are homonyms. Example: 'Viola' – in AGROVOC it is the taxonomic name of a plant (violets) while in SWD it refers to a musical instrument. In this case the relationship is 0. Sense disambiguation techniques such as the ontology partitioning performed by some of the current mapping systems, like Falcon-AO, should be able to solve most of these ambiguities by recognizing that none of the broader or narrower terms of 'Viola' and 'violet' are similar.

Some of the mappings of course will remain impossible for automatic methods that do not exploit sources of background knowledge, for example one of the AGROVOC-SWD mappings that found that 'Kater' (tomcat) is a 'männliches Individuum' (male individual).

## 6. Conclusion

The current mappings in the project at GESIS-IZ will be further analyzed and leveraged for distributed search not only in the sowiport portal but also in the German interdisciplinary science portal vascoda. Some of these mappings are already in use for the domain-specific track at the CLEF (Cross-Language Evaluation Forum) retrieval conference. We also plan on leveraging the mappings for vocabulary help in the initial query formulation process as well as for the ranking of retrieval results (Mayr, Mutschke & Petras, 2008).

We have seen that automatic mapping can definitely be very helpful and effective in case of Simple and Easy Lexical mappings. From our results, it appears that groups like Taxonomic vocabulary, Biological and Chemical Terminology and Geographic concepts fall into this category, as in general there seems to be more consensus on how to name things than in other groups. However, we need to be careful in these areas, where often word similarity does not mean that this is a potential mapping. These can be serious traps for automatic mapping approaches (like in the case of geopolitical issues).

Things get potentially more difficult in the case of more diversified groups/categories (in our case just summarized as Miscellaneous). Here, often background knowledge is needed to infer the correct mapping, and automatic mapping tools are able to identify only very little of these correctly. Most of the automatic suggestions are simply wrong or should not be equivalence relationships but broader, narrower or related terms.

---

[15] http://de.wiktionary.org/wiki/Falke or http://de.wiktionary.org/wiki/Greifvogel.

The bottom line is that for the moment, mapping should not be seen as a monolithic exercise, but we can take the best of both approaches and use automatic mapping approaches to get to the simple and easy lexical mappings and then use human knowledge to control the ambiguous cases.

## Acknowledgments

We would like to thank Lori Finch at the NAL for her extensive help on the AGROVOC-NALT mapping and for many discussions that contributed to this work. Van Hage was supported by the Dutch BSIK project Virtual Laboratory for e-science (http://www.vl-e.nl). The project at GESIS-IZ was funded by the German Federal Ministry for Education and Research, grant no. 01C5953. P. Mayr wishes to thank all our project partners and my colleagues in Bonn for their collaboration.